\documentclass[11pt]{advances2-preprint} 

\newtheorem{theorem}{Theorem}[section]

\theoremstyle{definition}

\theoremstyle{remark}

\copyrightinfo{}
{}

\newcommand \Psib {\overline \Psi} 
\newcommand \Phib {\overline \Phi} 

\newcommand \ab {\overline a} 
 
\newcommand \bU {\underline U} 
\newcommand \bPhi {\underline \Phi} 
\newcommand \bPsi {\underline \Psi}  
\newcommand \ba {\underline a} 
  
\newcommand \bNcal {\underline {\mathcal N}} 
\newcommand \Ncalb {\overline {\mathcal N}}

\newcommand \Mt {\widetilde M} 
\newcommand \TT 	 	{\mathfrak{T}} 
\newcommand \XX 	 	{ \TT_0^1}
\newcommand \YY 	 	{ \TT_1^0}
\newcommand \dM 	 	{{m}}

\newcommand \rad 		{\succ_{{\mathcal D}',{\mathcal D}}}
\newcommand \lad 		{\prec}
\newcommand \Dirac 		{{\boldsymbol \delta}} 
\newcommand \Y   	 	{ T_1^0}
\newcommand \X   	 	{ T_0^1}
\newcommand \Hbar {\overline H}

\newcommand \gbf        g
\newcommand \Tbf        T

\newcommand \Dbf      \nabla
\newcommand \Tbft T

\newcommand \del   {{\partial}} 
\newcommand \eps   {\varepsilon}

\newcommand \be {\begin{equation}}
\newcommand \ee {\end{equation}}

\newcommand \Pcal   {\mathcal P}    
\newcommand \Hcal {\mathcal H}   
\newcommand \Mcal {\mathcal M} 
\newcommand \Lcal {\mathcal L}

\newcommand \Ccal {\mathcal C}

\newcommand \lam \lambda

\newcommand \ra    {\big\rangle}
\newcommand \la    {\big\langle} 
\newcommand {\half}{\tfrac{1}{2}}
   
\newcommand \HH    {{\mathcal H}}

\newcommand \Ric       {\text{\bf Ric}}

\newcommand \kbar {\overline k}

\newcommand \Vb    {\overline V}

\newcommand \Dcal {\mathcal D}  
\newcommand \Ncal {\mathcal N}  

\newcommand \loc {\text{loc}} 
\newcommand \Bcal 	{{\mathcal B}}

\newcommand \Riem   	{\textbf{Rm}}

\begin{document}
\setcounter{page}{1}

\title[Einstein Spacetimes with Weak Regularity]{Einstein Spacetimes with Weak Regularity\footnote{Published in 
``Advances in Lorentzian Geometry'', Ed. M. Plaue, A.D.  Rendall, and M. Scherfner, 
AMS/IP series, Vol. 49, 2011, pp.~81--96.} }

\author[Philippe G. L{\tiny e}Floch]{Philippe G. LeFloch}
\address{Laboratoire Jacques-Louis Lions \& Centre National de la Recherche Scientifique, 
Universit\'e Pierre et Marie Curie (Paris 6)
\\
4 Place Jussieu, 75252 Paris, France.
 {\sl Blog address: \tt philippelefloch.wordpress.com} 
}
              \email{pgLeFloch@gmail.com}

\date{September 2010}

\begin{abstract} We review recent work on the Einstein equations of general relativity 
when the curvature is defined in a weak sense. Weakly regular spacetimes are constructed, 
in which impulsive gravitational waves, as well as shock waves, propagate. 
\end{abstract}

\maketitle

\section{Introduction}

The Einstein field equations relating the Ricci tensor $Ric_g$, the scalar curvature $R_g$ and the energy-momentum tensor $T$ of a spacetime  $(M, g)$
\be
\label{EEE}
Ric_g - {R_g \over 2} g = 8\pi \, T 
\ee
can be imposed in a weak sense, provided the metric $g$ 
belongs to the Sobolev space $H^1$ of square-integrable functions 
(and satisfies a suitable $L^\infty$ condition); cf~\cite{LMardare} and the references therein. 
At this level of (weak) regularity, the spacetime may contain impulsive gravitational waves (propagating at the speed of light) as well as shock waves (propagating at about the speed of sound in the matter). 
While shock waves occur only when the Einstein equations are coupled to the Euler equations of compressible fluids, the weak regularity class is relevant even in vacuum spacetimes and allow for singular nonlinear wave patterns not observed in the regular class and yet relevant from 
the physical standpoint. 

In recent work by the author and collaborators \cite{BLSS,LStewart1,LMardare,LStewart2,LRendall,LSmulevici}, 
several classes of spacetimes were
constructed by solving the initial value problem for the Einstein equations 
when the initial data set has weak regularity and 
enjoys certain symmetry assumptions. 
This work encompasses especially spacetimes admitting two Killing fields
and satisfying the Einstein-Euler equations. 

All classical work on spacetimes with two Killing fields 
requires that data (and solutions) are sufficiently regular. 
When weak regularity, only, is assumed, many classical arguments and estimates 
are no longer valid, and new arguments of proof  
must be developed that are robust enough 
to establish the existence of weakly regular spacetimes
and determine their global geometric structure.
In comparison with classical results, by encompassing a larger class of 
initial data sets and, therefore, allowing for propagating waves of possibly singular nature, 
one need to go significantly beyond the techniques of analysis for regular spacetimes,
and one 
provides additional rigorous validation of 
the underlying physical theory (including Penrose conjecture) 
within a much broader context.

An outline of this review is as follows. In Section~\ref{sec2}, 
we consider general manifolds that need not admit symmetries, 
and we present a general definition of curvature for weakly regular metrics.  
In Section~\ref{sec3}, we focus on spacetimes with Gowdy symmetry 
when the matter is an isothermal perfect fluid, 
and we present a global foliation of the future 
development of an initial data set with weak regularity. 
Finally, in Section~\ref{sec4}, we consider the special class of 
plane-symmetric spacetimes when the matter is a null fluid, 
and we establish a version of Penrose's conjecture for the broad class of weakly regular spacetimes.


\section{Definition and properties of manifolds with weak regularity}
\label{sec2}

\subsection*{Distributions on manifolds}

We begin by presenting the definition of the curvature
 introduced in \cite{LMardare}, 
which is at the basis of our investigation. 
Our first task is to define distributions on an $m$-dimensional differential manifold $M$ (with smooth topological structure).  
The definition is motivated by the canonical embedding $f \in L^1_\loc(M) \mapsto f \in \Dcal'(M)$, 
determined by 
$$
\lad f,\omega\rad:=\int_M f\omega, \qquad \omega\in \Dcal\Lambda^\dM(M), 
$$ 
where $\Dcal\Lambda^m(M)$ denotes the space of 
all (compactly supported) $m$-form fields on $M$ (or densities, or by abuse of notation, volume forms). 
Precisely, the space of {\sl scalar distributions} $\Dcal'(M)$ is defined as the continuous dual of $\Dcal\Lambda^m(M)$. 
Analogously, the space of {\sl distribution densities} $\Dcal' \Lambda^m(M)$ is the continuous dual of the space 
$\Dcal(M)$ of all compactly supported functions on $M$.

Given a smooth vector field $X$ and an $m$-form field $\omega$, 
we denote by $i_X \omega$ the associated interior product. 
Given any open subset with regular boundary $\Omega \subset M$ and using 
the identity $\Lcal_X=di_X+i_Xd$ together with Stokes formula, we obtain  
$$
\int_{\Omega} (Xf) \, \omega = - \int_{\Omega} f \, \Lcal_X \omega  + \int_{\del \Omega} f \, i_X \omega.
$$ 
This observation motivates the following definition 
of the action $X f$ of a smooth vector field on a scalar distribution:  
\be
\label{001}
\lad Xf, \omega\rad := - \lad f,\Lcal_X\omega\rad,   \qquad \omega\in\Dcal\Lambda^\dM(M). 
\ee

The notion of tensor distributions is defined as follows. 
Denote by $\TT_q^p(M) := \Ccal^\infty T_q^p(M)$
the space of all smooth $(p,q)$-tensor fields on $M$, 
and define the space $\Dcal' T_q^p(M)$ of all {\sl $(p,q)$-tensor distributions} 
as the space of all $C^\infty(M)$-multi-linear maps 
$$
A : \underbrace{\XX(M)\times \ldots \times \XX(M)}_{q\text{ times}} \times 
        \underbrace{\YY(M) \times \ldots \times \YY(M)}_{p\text{ times}}\to \Dcal'(M). 
$$
For instance, the embedding $L^1_\loc T_q^p(M) \subset \Dcal' T_q^p(M)$ is canonically defined by 
$$
\aligned  
&   \lad A(X_{(1)},\ldots,X_{(q)},\theta^{(1)},\ldots,\theta^{(p)}),\omega\rad 
\\
&   := \int_M  A(X_{(1)},\ldots,X_{(q)},\theta^{(1)},\ldots,\theta^{(p)}) \, \omega 
\endaligned 
$$
for all $\omega \in \Dcal \Lambda^m(M)$, $X_{(1)},\ldots,X_{(q)} \in \XX(M)$ and 
$\theta^{(1)},\ldots,\theta^{(p)} \in \YY(M)$.

\subsection*{Connections on manifolds}

A{\sl distributional connection} $\nabla: \XX(M) \times \XX(M) \to \Dcal' \X(M)$, by definition, 
satisfies the standard linearity and Leibnitz properties for all smooth fields. 
However, this class is often too broad in the applications, and
 more regularity may be assumed. 

An {\sl $L^2_\loc$ connection,} by definition, has $\nabla_XY\in L^2_\loc\X(M)$ for all $X,Y\in\XX(M)$, 
and its canonical extension: $\nabla:\XX(M)\times L^2_\loc\X(M)\to \Dcal'\X(M)$  is defined by 
$$
\aligned 
&   \la\nabla_X Y,\theta\ra  = X(\la Y,\theta\ra) - \la Y,\nabla_X\theta \ra \quad \text{ in } \Dcal'(M),
\\
&   X\in\XX(M), \quad   Y \in L^2_\loc\X(M), \quad   \theta\in \Dcal\Y(M).  
\endaligned 
$$
In this class of connections, we define the {\sl distributional Riemann tensor}  
$$
\Riem:\XX(M)\times \XX(M)\times \XX(M) \to \Dcal'\X(M), 
$$
as an element of the dual space $H^{-1}_\loc$ by 
$$
\aligned 
\la \Riem(X,Y)Z,\theta\ra 
= 
& X\la\nabla_Y Z, \theta\ra -Y\la\nabla_X Z, \theta\ra
\\
& -\la\nabla_Y Z, \nabla_X\theta\ra+\la\nabla_X Z, \nabla_Y\theta\ra-\la\nabla_{[X,Y]} Z, \theta\ra 
\endaligned
$$
for all $\theta\in \YY(M)$, $X,Y,Z \in \XX(M)$. By taking the trace in the usual manner, 
we also define the {\sl distributional Ricci tensor} $\Ric$ in $H^{-1}_\loc$. 
 
\subsection*{Sequences of connections}

One can establish the following stability property: if 
$\nabla^{(n)}$ is a sequence of $L^2_\loc$ connections on $M$, 
converging in $L^2_\loc$ to some connection $\nabla^{(\infty)}$, i.e. 
$$
\nabla^{(n)}_XY\to \nabla^{(\infty)}_X Y \qquad \text{ in } L^2_\loc
$$
for all $X,Y \in \XX(M)$, 
then the distributional Riemann and Ricci tensors 
$\Riem^{(n)}$ and $\Ric^{(n)}$ of the connections converge
to the distributional curvature tensors  
of the limit, i.e.
$$
\Riem^{(n)} \to \Riem^{(\infty)}, \qquad \Ric^{(n)} \to \Ric^{(\infty)} 
$$
in the distribution sense.
   

\subsection*{Distributional Levi-Cevita connections}

Consider next the general class of distributional metrics. Observe that the compatibility condition
$\nabla g =0$ between a metric and its associated Levi-Cevita connection
must be handled with care,
since non-smooth connections do not act on non-smooth tensors. 

The {\sl distributional Levi-Cevita connection} of a distributional metric $g$ is the operator 
$\nabla^\flat:(X,Y)\in\XX(M)\times\XX(M)\mapsto \nabla^\flat_X Y\in\Dcal' \Y(M)$, 
defined by the {\sl ``dual'' Koszul formula}
\be
\label{002}
\aligned
\la \nabla^\flat_X Y,Z \ra:=\frac12 \Big(
& X(g(Y,Z))+Y(g(X,Z))-Z(g(X,Y))\\
& -g(X,[Y,Z])-g(Y,[X,Z])+g(Z,[X,Y])\Big). 
\endaligned
\ee
It follows that in the weak sense 
$$
\aligned
& \nabla^\flat_X Y-\nabla^\flat_Y X - [X,Y]^\flat=0,
\\
& X(g(Y,Z))-\la\nabla^\flat_X Y,Z\ra-\la Y,\nabla^\flat_X Z\ra=0
\endaligned
$$
for all $X,Y,Z\in\XX(M)$.

One can check the following stability property under distributional convergence. 
Let $g^{(n)}$ be a sequence of distributional metrics converging in $\Dcal'$ to some limiting metric $g^{(\infty)}$. 
Then, the distributional connection ${\nabla^\flat}^{(n)}$ associated with $g^{(n)}$
converges in $\Dcal'$ to the connection ${\nabla^\flat}^{(\infty)}$ 
associated with $g^{(\infty)}$, i.e., for all $X,Y,Z\in\XX(M)$
$$
\la {\nabla^\flat}^{(n)}_X Y, Z \ra \to \la {\nabla^\flat}^{(\infty)}_X Y, Z \ra 
\quad 
\text{ in }\Dcal'(M).
$$

\subsection*{Distributional curvature of metrics}

In practice, further regularity is often assumed and, especially, the class of $H^1$ regular 
metrics plays an important role. 

If $\nabla$ is the Levi-Cevita connection of a metric $g$ of class $H^1_\loc\cap L^\infty_\loc$ which 
is also uniformly non-degenerate (for some $c>0$)
\be
\label{cond77}
|\det (g)| \geq c,  
\ee
then it is of class $L^2_\loc$ and its Riemann and Ricci curvature tensors are well-defined as distributions. 
Moreover, its scalar curvature $R$ is also well-defined as a distribution: 
$$
R = g^{\alpha\beta} \Ric(E_{(\alpha)},E_{(\beta)}),
$$
where $E_{(\alpha)}$ is a local frame.

Namely, since $H^1_\loc\cap L^\infty_\loc$ is an algebra and $g$ is 
uniformly non-degenerate, it follows that $g^{\alpha\beta}\in H^1_\loc\cap L^\infty_\loc(M)$. On the other hand, one can write 
$$
\aligned
\Ric(E_{(\alpha)},E_{(\beta)})
&= \la E^{(\sigma)}, \Riem(E_{(\alpha)},E_{(\sigma)})E_{(\beta)}\ra\\
& 
   = E_{(\alpha)}(\la E^{(\sigma)}, \nabla_{E_{(\sigma)}}E_{(\beta)}\ra )
  - E_{(\sigma)}(\la E^{(\sigma)}, \nabla_{E_{(\alpha)}}E_{(\beta)}\ra )\\
& \quad 
- \la \nabla_{E_{(\alpha)}}E^{(\sigma)}, \nabla_{E_{(\sigma)}}E_{(\beta)}\ra
+ \la \nabla_{E_{(\sigma)}}E^{(\sigma)}, \nabla_{E_{(\alpha)}}E_{(\beta)}\ra
\\
& \quad - \la E^{(\sigma)}, \nabla_{[E_{(\alpha)},E_{(\sigma)}]}E_{(\beta)}\ra.
\endaligned
$$
Since the last three terms belong to $L^1_\loc(M)$, we only need to define the product 
of $g^{\alpha\beta}$ and the distributions 
$$
E_{(\alpha)}(\la E^{(\sigma)}, \nabla_{E_{(\sigma)}}E_{(\beta)}\ra ), 
\qquad 
E_{(\sigma)}(\la E^{(\sigma)}, \nabla_{E_{(\alpha)}}E_{(\beta)}\ra ).
$$
This is done by letting (for the first term, say, since the second term is similar) 
$$
\aligned
& g^{\alpha\beta}E_{(\alpha)}(\la E^{(\sigma)}, \nabla_{E_{(\sigma)}}E_{(\beta)}\ra )
\\
& :=E_{(\alpha)}(g^{\alpha\beta} \la E^{(\sigma)}, \nabla_{E_{(\sigma)}}E_{(\beta)}\ra )
- (E_{(\alpha)}g^{\alpha\beta})\la E^{(\sigma)}, \nabla_{E_{(\sigma)}}E_{(\beta)}\ra,
\endaligned
$$ 
which, clearly, is well-defined as a distribution.

\subsection*{Sequences of metrics}

One can check the following property of stability under strong sconvergence. 
Let $g^{(n)}$ be a sequence of $H^1_\loc$ metric tensors converging locally 
in $H^1_\loc$ to some limiting metric $g^{(\infty)}$, and 
assume that the inverse metrics $g^{-1}_{(n)}$ converge locally in $L^\infty_\loc$ 
to $g^{-1}_{(\infty)}$:  

\begin{enumerate}

\item[$\bullet$] The Levi-Cevita connections ${\nabla}^{(n)}$ associated with $g^{(n)}$ are of class $L^2_\loc$ 
and converge in $L^2_\loc$ to the connection ${\nabla}^{(\infty)}$ of 
the limit $g^{(\infty)}$, that is, for all $X,Y \in \XX(M)$,  
$$
{\nabla}^{(n)}_XY\to {\nabla}^{(\infty)}_X Y \qquad \text{ strongly in } L^2_\loc.
$$

\item[$\bullet$] The distributional Riemann, Ricci, and scalar curvature tensors $\Riem^{(n)}$, $\Ric^{(n)}$, 
$R^{(n)}$ of the connections $\nabla^{(n)}$ converge in $\Dcal'$
to the limiting curvature tensors $\Riem^{(\infty)}$, $\Ric^{(\infty)}$, $R^{(\infty)}$ of $\nabla^{(\infty)}$, respectively. 
\end{enumerate}
 
Hence, the curvature of a spacetime may be defined in the sense of distributions
when the metric has weak regularity, only.  
We have advocated here the use of fully geometric definitions, 
while coordinate dependent definitions are usually 
adopted in the literature; see Geroch and Traschen \cite{GT}, Mars and Senovilla \cite{MS}, and Lichnerowicz \cite{Lichne}.  Other approaches are developed for certain classes of spacetimes
by Kunzinger, Steinbauer, Vickers, Balasin, Aichelburg, Heinzle, etc. 

A framework to deal with manifolds with low regularity was also introduced 
by Lott, Villani, Sturm, 
and allows to define Ricci lower bounds via the theory of optimal transport. 
Both lower and upper curvature bounds can also be defined in the theory of Alexandrov spaces. 


\subsection*{Discontinuous connections.}

Consider connections having a jump discontinuity 
across a given smooth hypersurface $\HH\subset M$. 
A {\sl connection that is discontinuous across $\Hcal$} is, by definition, 
a connection $\nabla$ of class $L^2_\loc(M) \cap W_\loc^{1,p}(M^\pm)$ 
having a jump discontinuity across $\Hcal$, with 
$$
M = M^- \cup M^+, \qquad M^- \cap M^+ = \HH. 
$$
Clearly, according to our definitions above, the corresponding operators $\nabla^\pm$ determined from $\nabla$ by restriction to 
each $M^\pm$ 
have well-defined, distributional Riemann and Ricci curvatures $\Riem^\pm$ and $\Ric^\pm$.  
In fact, these curvature tensors belong to $L^1_\loc(M^\pm)$, at least, and actually to $L^p_\loc(M^\pm)$ if $p \geq m/2$. 

Our purpose now is to compute the distributional curvature of the connection $\nabla$, which 
we expect to contain Dirac mass singularities along $\HH$.

First of all, given a smooth hypersurface $\Hcal$, 
the {\sl Dirac measure supported by $\HH\subset M$} 
is defined as the $1$-form distribution  $\Dirac_{\HH} \in \Dcal'\Y(M)$ such that 
$$
\aligned 
&   X \in \XX(M)\mapsto \la \Dirac_{\HH},X\ra\in \Dcal'(M), 
\\
&   \lad \la \Dirac_{\HH}, X \ra, \omega \rad = \int_\HH i_X \omega, \qquad \omega \in \Dcal \Lambda^\dM(M). 
\endaligned 
$$ 
We write $[ A ]_\HH:= A^+ - A^-$ for the jump of a tensor field $A$ across $\HH$, 
and we write the ``regular part'' as 
$$
A^{reg}:=
\begin{cases}
A^+ & \text{ in } M^+,
\\
A^- & \text{ in } M^-. 
\end{cases}
$$ 
 
For instance, the distributional derivative $\nabla V$ of a vector field 
(smooth in $M^\pm$ and discontinuous across $\HH$) is  
\be
\label{003}
\nabla_X V:=(\nabla_X V)^{reg} +[V]_\HH \, \la \Dirac_{\HH},X\ra
\ee
for all $X \in \XX(M)$. 
The term $\la \Dirac_\HH,X\ra$ depends on $X$ only via its restriction to $\HH$, so 
$\Dirac_\HH$ can be applied to vector fields only defined on $\HH$. 
If $X$ is a vector field tangent to the hypersurface ($X \in \TT^1_0(\HH)$), 
then $\la\Dirac_{\HH},X\ra=0$.

\subsection*{Jump relations} 

We can now derive jump relations associated with a discontinuous connection. Choose an adapted frame
so that $E_{(i)}$, $i=1, \ldots, m-1$ is a local frame on the hypersurface $\HH$, while 
$E_{(\alpha)}$, $\alpha= 1,\ldots,m$: is a local frame on $M$, and 
let $E^{(\alpha)}$: be the associated dual frame of $1$-form fields. 
Then, the distributional Riemann curvature takes the form 
$$
\aligned   
\Riem(X,Y)Z = (\Riem(X,Y)Z)^{reg}
&   + [\nabla_Y Z ]_\HH \la \Dirac_\HH, X\ra 
\\
&   - [\nabla_X Z]_\HH \la\Dirac_\HH, Y\ra 
\endaligned 
$$ 
for all $X,Y,Z \in \XX(M)$, while the distributional Ricci curvature reads 
$$
\aligned   
\Ric(X,Y) = (\Ric(X,Y))^{reg} 
&   + [\la E^{(\alpha)},\nabla_{E_{(\alpha)}}Y \ra]_\HH \la\Dirac_\HH, X\ra 
\\
&   - [\la E^{(m)},\nabla_X Y \ra]_\HH \la\Dirac_\HH, E_{(m)}\ra  
\endaligned 
$$  
for all $X,Y \in \XX(M)$. 

From these formulas, we can deduce some properties of the jumps.  
Consider a connection $\nabla$ that is discontinuous across a hypersurface $\Hcal$.

\begin{itemize}

\item The singular part of the Riemann tensor vanishes if and only if 
the connection $\nabla$ is continuous across $\HH$.  
 
\item The singular part of the Ricci tensor vanishes if and only if the components 
$$
\aligned 
&   \la E^{(m)},\nabla_{E^{j)}} X \ra \, \text{ and } 
  \la E^{(j)},\nabla_{E_{(j)}}X \ra \, \text{ are continuous across }   \HH
\\
&  \text{for all vector field $X\in \XX(M)$.}
\endaligned
$$
  
\item Suppose $\nabla$ is the $L^2$ Levi-Cevita connection of a uniformly non-degenerate metric $g$
and suppose that $g$ is $W_\loc^{2,p}$ on each side of some hypersurface $\Hcal$, then
$$
R := R^{reg}+[\la g^{m\beta}E^{(j)}-g^{j\beta} E^{(m)},\nabla_{E_{(j)}}E_{(\beta)}\ra]_\HH \la\Dirac_\HH, E_{(m)}\ra. 
$$
where $E^{(j)}$ is a frame adapted to $\Hcal$.

\end{itemize}  

 
\section{A global foliation for Einstein-Euler spacetimes with Gowdy-symmetry on $T^3$} 
\label{sec3}

\subsection*{Spacetimes with Gowdy symmetry}
  
This section is based on the paper \cite{LRendall} in collaboration with A.D. Rendall.    
We impose that the spacetime $(M,g)$ under consideration has Gowdy symmetry, which 
is a classical assumption made to study inhomogeneous cosmology with a ``big bang'' or ``big crunch''.
As far as the matter model is concerned, 
we consider a compressible, isothermal perfect fluid
 described by the energy-momentum tensor
\be
\label{020}
T^{\alpha\beta} = (\mu + p) u^\alpha u^\beta + p \, g^{\alpha\beta}
\ee 
with future oriented, timelike velocity vector $u^\alpha$, normalized so that 
$$
g_{\alpha\beta} u^\alpha u^\beta=-1,
$$ 
and with mass-energy density $\mu \geq 0$ and pressure $p = c_s^2 \, \mu$. 
Here, $c_s \in (0,1)$ does not exceed the light speed normalized to be $1$. Here, all greek indices $\alpha, \beta,\ldots$ 
describe $0,1,2,3$.

Our main objective here is to construct a class of
 globally foliated spacetimes, obtained by solving the initial value problem  
when initial data have low regularity. Under our construction, 
gravitational waves (possibly impulsive ones)  at the speed of light 
can propagate in such spacetimes, 
and shock waves can also propagate at (about) the speed of sound.

Gowdy symmetry on $T^3$ imposes that the $(3+1)$-dimensional Lorentzian manifold $(M,g)$
admits the Lie group $T^2$ as an isometry group acting on the torus $T^3$: 
it is generated by two 
(linearly independent, orthogonally transitive) spacelike vector fields 
$X,Y$ satisfying, by definition, the Killing condition 
$$
\Lcal_X g = \Lcal_Y g = 0, 
$$
the commutation condition 
$$
[X,Y] = 0,
$$ 
the spacelike property 
$$
g(X,X), \, g(Y,Y) >0. 
$$
In addition, by definition, 
we impose the property of orthogonal transitivity, 
in the sense that 
the distribution of $2$-planes of covectors 
$$
\text{Vect}(X,Y)^\perp
$$  
is Frobenius integrable. 

In the vacuum, the above conditions determine exactly the so-called {\sl Gowdy spacetimes.}
Note that the above property of integrability can be analyzed as follows. 
By introducing two vectors $Z,T$ orthogonal to $X,Y$, we can ensure
 that (using a
reduction to dimension $2$ and after normalization)
$$
\eps(X,Y, \cdot, \cdot)^\sharp = Z \otimes T - T \otimes Z, 
$$
where $\eps$ is the canonical volume form, or equivalently
$$
\eps^{\alpha\beta\gamma\delta} X_\gamma Y_\delta = Z^\alpha T^\beta - T^\alpha Z^\beta. 
$$ 
Then, the distribution of covectors $g(X, \cdot), \, g(Y, \cdot)$ is Frobenius integrable if and only if 
$Z,T, [Z,T]$ are linearly dependent,
that is, $\eps(Z,T, [Z,T], \cdot) = 0$ or 
$$
\eps_{\alpha\beta\gamma\delta} Z^\alpha T^\beta \big[Z,T]^\gamma = 0. 
$$
After some calculations, this condition is found to be equivalent to vanishing 
``twist constants'' (as identified by Chrusciel \cite{Chrusciel}) 
$$
\la \eps(X,Y, \cdot, \cdot), \nabla X \ra_g = \la \eps(X,Y, \cdot, \cdot), \nabla Y \ra_g = 0, 
$$
i.e. 
$$  
\eps_{\alpha\beta\gamma\delta} X^\alpha Y^\beta \nabla^\gamma X^\delta
=
\eps_{\alpha\beta\gamma\delta} X^\alpha Y^\beta \nabla^\gamma Y^\delta
= 0. 
$$ 
Note that the vacuum Einstein equations imply that the above (twist) quantities are constants, while this is an assumption in matter spacetimes.

\subsection*{Main existence result} 
 
A {\sl Gowdy symmetric initial data set on $T^3$} 
$$
\big(\overline H, 
\overline g, \overline k, \overline \rho, \overline j \big)
$$
 is determined by 
a Riemannian $3$-manifold $(\Hbar,\overline g)$
together with a symmetric tensor field $\kbar$ (playing the role of the second fundamental form of the spacetime
to be constructed), in which
$\overline\rho$ denotes the energy density of the fluid (a scalar field) and $\overline j$ 
its momentum (a vector field, essentially equivalent to the velocity field) 
defined on $\Hbar$.  In addition, the so-called Einstein constraint equations must 
hold, and these data must be invariant under Gowdy symmetry. These initial data is assumed to be solely weakly regular. 
 
The class of {\sl weakly regular spacetimes with Gowdy symmetry on $T^3$} is defined
by the following conditions:

\begin{itemize}

\item The metric belong to the Sobolev space  $H^1(\Sigma)$ on every spacelike slices $\Sigma \subset M$. 
     
\item The mass-energy density $\rho \geq 0$ (scalar field) and the momentum $j$ (vector field) 
belongs to the Lebesgue space $L^1(\Sigma)$ of integrable functions, 
for every spacelike slice $\Sigma$. (Prescribing $\rho, j$ is equivalent to prescribing $\mu, u$.)  
 
\item The Einstein equations \eqref{EEE} are satisfied in the distribution sense and, in particular, 
the curvature is well-defined as a distribution in $H^{-1}(\Sigma)$ for each slice $\Sigma$.   

\item Furthermore, the entropy inequalities associated with the Euler equations are satisfied.  

\end{itemize}

\begin{theorem}[Einstein-Euler spacetimes with Gowdy symmetry on $T^3$] 
\label{maintheo} 
Let $\big(\overline g, \overline k, \overline \rho, \overline j \big)$ be a 
weakly regular, Gowdy symmetric, initial data set on $T^3$
for the Einstein-Euler equations, and assume that these initial data have constant area,  
and are everywhere expanding or 
everywhere contracting. Then, there exists a weakly regular, 
Gowdy symmetric spacetime $(M, g, \rho, j)$
satisfying the Einstein-Euler equations  
in the distributional sense, so that the following properties hold.
The manifold $(M,g,\rho, j)$ is (up to diffeomorphisms) a Gowdy-symmetric 
future development of 
$\big(\overline g, \overline k, \overline \rho, \overline j \big)$   
which is globally covered by a single chart of coordinates $t$ and $(\theta,x,y) \in T^3$, with 
$$
M = 
\begin{cases}
\quad \big\{ (t, \theta) \, / \, 0 < c_0 \leq t < \infty \big\} \times T^2,   &\text{ expanding case,} 
\\
\\ 
\quad \big\{ (t, \theta) \, / \, c_0 \leq t < c_1 \leq 0 \big\} \times T^2,   &\text{ contracting case.} 
\end{cases}
$$ 
Here, $c_1 \in (c_0,0]$ is a constant, and the time variable is chosen to coincide  
with the area of the surface of symmetry in the expanding case,
 and with minus this area in the contracting case.  
\end{theorem}

\subsection*{Further comments and perspectives}

Many earlier works on {\sl regular}
spacetimes with Gowdy symmetry are available, both, in the vacuum case
(Berger, Chrusciel, Isenberg, Moncrief, Weaver) and 
when the Einstein equations are coupled to a kinetic matter models
(Rendall, Andreasson, Dafermos); 
cf.~\cite{Moncrief} and \cite{LRendall} for references.
For an analysis of the global geometry 
of $T^2$--symmetric vacuum spacetimes with weak regularity, we refer to LeFloch and Smulevici 
\cite{LSmulevici}, and 
for an earlier result on compressible fluid models
(a ``small time'' result for spacetimes with bounded variation)
 to LeFloch and Stewart \cite{LStewart2} (see also \cite{BLSS}).

Note that the time-function constructed above is geometric in nature and 
is based on the Gowdy symmetry assumption (areal coordinates). It would be interesting to investigate whether a 
foliation by constant mean curvature slices also exists.  
Even more important is to investigate the structure of the boundary of the future development of the constructed spacetimes.   
In general, the constant $c_1$ need not vanish, since 
explicit counter-examples for the Einstein-Euler equations are available, although such examples
are probably unstable.

On the other hand, 
recall that spacetimes with bounded variation were constructed by 
Chris\-to\-doulou in his work \cite{Christo1,Christo2} that settled positively 
the weak version of Penrose's 
cosmic censorship conjecture in the context of spherically symmetric spacetimes and for scalar fields. Recall also that Groah and Temple \cite{GroahTemple}
established a {\sl local-in-time} existence result for spherically symmetric matter spacetimes. 
In such spacetimes, no gravitational waves are permitted and the matter equations are coupled with a 
differential equation accounting for (non-evolutive) geometrical features.

Following Gowdy and Geroch, one considers the 
quotient manifold $\Mt:= M/T^2$. Since the two Killing fields 
$(X,Y)=:(X_1, X_2)$ are linearly independent and spacelike, the area of the orbit of 
orbits of the $T^2$ symmetry group
satisfies 
\be
\label{030}
R^2:= \det \Big(\lambda_{ab}\Big) >0, \qquad \lambda_{ab} := g(X_a, X_b).  
\ee
Due to the Killing field property, $R>0$ is a constant on each orbit. 

In addition, the associated Lorentzian metric defined on $\Mt$ can be expressed 
as $h := g - \lambda^{ab} \, X_a \otimes X_b$.    
and one can also introduce 
the projection operator:
$$
h_\mu^\nu:= g_\mu^\nu - \lambda^{ab} X_{a\mu} X_b^\nu,
$$
which serves to project the Einstein equations on the quotient manifold. 

This leads us to introduce a set of local coordinates $(t, \theta, x, y)$ such that 
$x,y$ span the orbits of symmetry, and 
the metric coefficients depend on $(t,\theta)$, only, and are periodic in $\theta$. 
A key property that allows to choose the area as a time-function is that
the gradient $\nabla R$ is timelike ---a fact that remains true for the Euler equations and even in the weak regularity class
under consideration. 

More precisely, in suitably {\sl conformal coordinates} $(\tau,\theta,x,y)$ one can arrange that the metric reads 
\be
\label{040}
g = e^{2(\eta-U)} \, (- d\tau^2 + d\theta^2) + e^{2U} (dx + A \, dy)^2 + e^{-2U} \, R^2 \, dy^2, 
\ee
where $\eta, U, A, R$ with $R>0$ depend on $\tau, \theta$, and 
the area $R$ is one of the unknowns. 
The Einstein equations reduce to a coupled system of eight partial differential equations, consisting 
of four evolution equations (second-order, semi-linear wave equations for $U, A, \eta, R$) and 
two constraint equations (nonlinear differential equations for $\nabla R$), which also imply 
the Euler equations (nonlinear hyperbolic equations for $\rho, j$, or $\mu, u$).

In {\sl areal coordinates} $(t,\theta,x,y)$ one has
\be
\label{050}
g = e^{2(\nu-U)} \, (- dt^2 + \alpha^{-1} d\theta^2) + e^{2U} (dx + A \, dy)^2 + e^{-2U} \, t^2 \, dy^2, 
\ee
in which $U,A, \nu, \alpha$ depend on $t, \theta$,  
and the time $t=R$ coincides with the area of the orbits of symmetry. The Einstein equations 
reduce to a coupled system of six partial differential equations, specifically 
three evolution equations (second-order, nonlinear wave equations for $U, A, \nu$) and   
 three constraint equations (including 
 nonlinear differential equations for $\alpha$), while the  
 Euler equations again form a system of two nonlinear hyperbolic equations. 
 

\section{The characteristic initial value problem for 
plane symmetric spacetimes with weak regularity} 
\label{sec4}

\subsection*{Objective in this section} 

Based on the paper \cite{LStewart2} in collaboration with J.M. Stewart,
we consider the class of spacetimes that satisfy the ``polarized'' Gowdy symmetry property
(hypersurface orthogonal property), and we investigate the Einstein equations when the matter 
is a null perfect fluid (cf.~\eqref{333}, below). 

We are interested here in the characteristic initial value problem when
data are prescribed on two null hypersurfaces, denoted by 
$\Ncal_{u_0}$ and $\Ncal_{v_0}$, intersecting along a $2$-plane $\Pcal_{u_0, v_0}$. 
Like in the previous section, spacetimes with weak regularity are constructed by solving the initial value problem. 
Our set-up and assumptions allow us to tackle a new issue, beyond the sole existence of the spacetimes, 
that is, to establish a suitable version of Penrose's strong censorship conjecture in this class.

Recall that the characteristic initial value problem 
 addressed here 
 for matter spacetimes with symmetry and weak regularity,  
was treated earlier within the class of sufficiently regular vacuum spacetimes
 by 
Friedrich~\cite{Friedrich}, Stewart and Friedrich \cite{SF},  Rendall~\cite{Rendall}, 
Christodoulou \cite{Christodoulou5}, Dossa
and Tadmon \cite{Dossa},
and 
 Choquet-Bruhat, Chrusciel, and Mart\'{\i}n-Garc\'{\i}a~\cite{CCM2}.

\subsection*{Matter model of interest}

The matter model under consideration is a null, irrotational fluid, whose 
sound speed, by definition, coincides with the speed of light (normalized to unit). In this context, 
propagating discontinuities in the fluid variables may occur only along null hypersurfaces.  
More precisely, a perfect fluid with pressure $p=\mu$ equal to its mass-energy density $\mu$ (null fluid) 
has energy-momentum tensor
\be
\label{333}
T^{\alpha\beta}  
= 2 \mu \, u^\alpha u^\beta + \mu \, g^{\alpha\beta}.  
\ee
As usual, $u$ is the velocity vector, normalized so that $u^\alpha u_\alpha = -1$.

The (second) contracted Bianchi identity allow us to derive the Euler equations  
$$
  \nabla_\alpha T^{\alpha\beta} = 0, 
$$
which take the form  
\be
\label{444}
(u^\alpha \, \nabla_\alpha \mu) \, u^\beta + \mu \, (\nabla_\alpha u^\alpha) \, u^\beta  
+ \mu \, u^\alpha \, \nabla_\alpha u^\beta - {1 \over 2} \nabla^\beta \mu = 0. 
\ee

Multiplying \eqref{444} by $u_\beta$ we obtain the scalar equation  
$$
 2 \nabla_\alpha w \, u^\alpha - 2w \, \nabla_\alpha u^\alpha 
+ 2w \, u^\alpha u_\beta \nabla_\alpha u^\beta
 - \nabla_\alpha w \, u^\alpha = 0, 
$$
which, in view of $u_\beta \nabla_\alpha u^\beta = 0$,
simplifies 
$$
u^\alpha \nabla_\alpha w + 2w \, \nabla_\alpha u^\alpha = 0, 
$$
or equivalently, assuming that the density is bounded away from zero and setting 
$
\Sigma = \half  \log w, 
$ 
we obtain 
$$
u^\alpha \nabla_\alpha \Sigma + \nabla_\alpha u^\alpha = 0. 
$$

Next, multiply by the projection operator $H_{\beta\gamma} = g_{\beta\gamma} - u_\beta u_\gamma$   
and obtain the vector equation 
$$
  H^{\alpha\gamma} \nabla_\alpha w - 2w \, u^\alpha  \nabla_\alpha u^\gamma = 0, 
$$
or equivalently 
$$
  H^{\alpha\gamma} \Sigma_{,\alpha} - u^\alpha \nabla_\alpha u^\gamma = 0. 
$$

In addition, we impose that the flow is irrotational, that is, we 
assume the existence of a (scalar) potential $\psi$ with timelike gradient  
$$
\nabla_\beta \psi \, \nabla^\beta \psi < 0, \qquad  
u^\alpha = \frac{\nabla^\alpha\psi}{\sqrt{|\nabla_\beta \psi \, \nabla^\beta|}}. 
$$ 
After normalization (by replacing $\psi$ with $F(\psi)$), the Euler equations imply the so-called 
Bernouilli's equation (in characteristic variables, cf. below) 
$$  
\mu = 
|\nabla^\alpha \psi \nabla_\alpha \psi| = 4 e^{-2a} \, \psi_u \psi_v, 
$$
which yields the energy density in terms of the potential.  
We arrive at a single matter equation for the scalar field $\psi$. 
In characteristic coordinates, this is a wave equation for the velocity potential 
$$
\psi_{uv} + b_v \, \psi_u + b_u \psi_v = 0. 
$$

\subsection*{Reduced form of the field equations}
 
Each Killing field is assumed to be hypersurface orthogonal (``plane symmetry'')
and we arrange the coordinates so that the metric reads 
$$
  \begin{aligned} 
    g
    & = e^{2a} \, (- dt^2 + dx^2) + e^{2b} \, (e^{2c} \, dy^2 + e^{-2c} \, dz^2)
    \\
    & = - e^{2a} \, du dv + e^{2b} \, (e^{2c} \, dy^2 + e^{-2c} \, dz^2), 
  \end{aligned}
$$
where the coefficients $a,b,c$ depend upon the characteristic variables 
$$
u= t-x, \qquad v=t+x, 
$$ 
only. The relevant components of the Einstein tensor are 
$$
  \begin{aligned} 
    G_{00} & = 2 \, \bigl( - 2 \, a_u b_u + b_{uu} + b_u^2 + c_u^2 \bigr), 
    \\
    G_{01} & = 2 \, \bigl( - b_{uv} - 2 \, b_u b_v \bigr),
    \\
    G_{11} & = 2 \, \bigl( -2 \, a_v b_v + b_{vv} + b_v^2 + c_v^2 \bigr),
    \\
    G_{22} & = 4 e^{-2a+2b+2c} \, \bigl( a_{uv} + b_{uv} + b_u b_v - b_u
    c_v - b_v c_u - c_{uv} + c_u c_v \bigr),
    \\ 
    G_{33} & = 4 e^{-2a+2b-2c} \, \bigl( a_{uv} + b_{uv} + b_u b_v + b_u
    c_v + b_v c_u + c_{uv} + c_u c_v \bigr). 
  \end{aligned} 
$$

We use here one of our gauge freedom, and solve a decoupled equation for $b$, 
allowing us to choose coordinates so that 
$$
e^{2b} = \half |u+v|,
$$
and the region of interest is 
$$
\bigl\{ u+v < 0 \bigr\}, 
$$ 
the hypersurface $u+v=0$ being a (possibly physical or coordinate) singularity.  

The corresponding evolution equations are two 
singular wave equations of Euler-Poisson-Darboux type 
\be
\label{777}
  \begin{aligned} 
    & \psi_{uv} + {1 \over 2 (u+v)} \, \big( \psi_u + \psi_v \big) = 0, 
    \\
    & c_{uv} + {1 \over  2 (u+v)} \, \big( c_u + c_v \big) = 0, 
  \end{aligned} 
\ee
while the constraint equations read 
\be
\label{778}
  \begin{aligned} 
    & a_u = (c_u^2 + \half  \psi_u^2) \, (u+v) - {1 \over 4 (u+v)},
    \\
    & a_v = (c_v^2 + \half  \psi_v^2) \, (u+v) - {1 \over 4(u + v)}.   
  \end{aligned} 
\ee

{\bf Characteristic initial value problem.}  

We can establish a global and fully geometric result, based on   
prescribing initial data on two null hypersurfaces intersecting along a two-plane. 
We seek for the global structure of the future development of a given initial data set. 
We consider arbitrary null coordinates, denoted below by $(U,V)$ 
and, throughout, we restrict attention to {\sl plane symmetric} data and spacetimes. 

An {\rm initial data with weak regularity}
consists of the following prescribed data.  
Let 
$$
\big(\bNcal, e^{\ba} dU dydz\big), \qquad \big(\Ncalb, e^{\ab} dV dydz\big)
$$ 
be two plane symmetric $3$-manifolds (endowed with volume forms) 
with boundaries identified along a two-plane $\Pcal$ and 
parametrized for some $(U_0, V_0)$ as  
$$
\bNcal := \big\{    U > U_0     \big\}, \qquad 
\Ncalb := \big\{    V > V_0     \big\}, 
\qquad   \Pcal := \left\{ U = U_0, \, V = V_0 \right\}.
$$ 
\begin{enumerate} 

\item Suppose that $\ba, \ab$ belong to the Sobolev space $W^{1,1}$, i.e.~the integrals 
$$
\int_{\bNcal} \big( |\ba| + |\del_U \ba| \big) \,  e^{\ba} \, dU
\qquad 
\int_{\Ncalb} \big( |\ab| + |\del_V \ab| \big) \,  e^{\ab} \, dV
$$ 
are finite, and are normalized so that $\ba|_\Pcal = \ab|_\Pcal =0$. 
\item Let $\bPsi_0, \bPhi_{00}$ and $\Psib_4, \Phib_{22}$ be (plane-symmetric) functions 
defined on the hypersurfaces  $\bNcal$ and $\Ncalb$, respectively, with  
$0 \leq \bPhi_{00} \in L^1(\bNcal)$ and $0 \leq \Phib_{22} \in L^1(\Ncalb)$, i.e.~the integrals 
$$
\int_{\bNcal} \bPhi_{00} \,  e^{\ba} \, dU, 
\qquad 
\int_{\Ncalb} \Phib_{22} \,  e^{\ab} \, dV
$$ 
are finite, 
and that 
$\bPsi_0 \in W^{-1,2}(\bNcal)$ and $\Psib_4 \in W^{-1,2}(\Ncalb)$, i.e. $\bPsi_0 = \del_U \bPsi_0^{(1)}$
and $\Psib_4 = \del_V \Psib_4^{(1)}$
with 
$$
\int_{\bNcal} \big|\bPhi_{00}^{(1)}\big|^2 \,  e^{\ba} \, dU, 
\qquad 
\int_{\Ncalb} \Big| \Phib_{22}^{(1)}\big|^2 \,  e^{\ab} \, dV. 
$$ 

\item Finally, one also prescribes the connection NP scalars $\rho_0, \sigma_0, \lambda_0, \mu_0$ on $\Pcal$. 
\end{enumerate}  

We then have the following existence and qualitative property result, for which
we refer the reader to \cite{LStewart2} about the terminology used here.

\begin{theorem}[Global causal structure of plane symmetric matter spacetimes]
\label{F-global} 
Consider an initial data set with weak regularity determined by
$\big(\bNcal, e^{\ba} \big)$ and $\big(\Ncalb, e^{\ab} \big)$, 
a plane $(\Pcal, \rho_0, \sigma_0, \lambda_0, \mu_0)$, 
and 
prescribed Ricci and Weyl NP scalars $\big(\bPsi_0, \bPhi_{00}\big)$ 
and $\big(\Psib_4, \Phib_{22}\big)$. 

(1) Then, there exists a $W^{1,2}$ regular spacetime $(\Mcal, g)$ determined by metric coefficients $a,b,c$
and matter potential $\psi$ which is a future development 
of the initial data set satisfying the Einstein equations  \eqref{EEE}  
for self-gravitating, irrotational fluids with the initial conditions
$$ 
(\rho, \sigma, \lambda, \mu) = (\rho_0, \sigma_0, \lambda_0, \mu_0) \quad \text{ on } \Pcal, 
$$
$$
\big(a, \Psi_0, \Phi_{00}\big) =  \big(\ba, \bPsi_0, \bPhi_{00}\big) \quad     \text{ on the null 
hypersurface } \bNcal,  
$$
and 
$$
\big( a, \Psi_4, \Phi_{22}\big) =  \big( \ab, \Psib_4, \Phib_{22}\big)  \quad \text{ on the null 
hypersurface } \Ncalb.  
$$  

(2) The constructed development of the initial data
has past boundary 
$$
\big\{  \bU_0 >  U > U_0; V=V_0    \big\} \cup \big\{ U=U_0; \Vb_0 > V > V_0     \big\} \subset \bNcal \cup \Ncalb
$$
and, for {\bf generic initial data} (in the sense defined in (3), below) the curvature blows up to 
(and makes no sense, even as a distribution) 
as one approaches  its ($W^{1,2}$ regular) future boundary
$$
\Bcal_0 := \big\{  F(U) + G(V) = 0     \big\}
$$ 
for some functions $F, G$, 
so that the spacetime is inextendible beyond $\Bcal_0$ within the class of $W^{1,2}$ regular spacetimes .

(3) The above result holds for generic initial data, in the sense that arbitrary data can always be perturbed in the natural 
(energy-type) norm so that the perturbed initial data do generate a singular spacetime whose curvature blows-up on 
$\Bcal_0$. 

\end{theorem}

The coefficients $e^{\ba}$ and $e^{\ab}$, modulo a conformal transformation, could 
be chosen to be identically $1$ on the initial hypersurface, so that 
two main degrees of freedom remain on each of the two initial hypersurfaces. 
Theorem~\ref{F-global} can be seen as a statement of Penrose strong's censorship conjecture 
for plane symmetric spacetimes with weak regularity.


\section*{Acknowledgments}

The author (PLF) was partially supported by the Centre National de la Recherche Scientifique (CNRS)
and the Agence Nationale de la Recherche (ANR) through the grant 06-2-134423 entitled
``Mathematical Methods in General Relativity'' (Math-GR).


\bibliographystyle{amsalpha}

\end{document}